\documentclass[pre,twocolumn,amsmath,amssymb]{revtex4}

\usepackage{graphicx}
\usepackage{latexsym}
\usepackage{amsmath}
\usepackage{amssymb}
\usepackage{amsfonts}
\usepackage{bm}

\newcommand{\la}{\left<}
\newcommand{\ra}{\right>}

\newcommand{\rvec}{\ensuremath{\underline{r}}}
\newcommand{\uvec}{\ensuremath{\underline{u}}}

\newcommand{\kBT}{\mbox{$k_{\rm B}T$}}

\newcommand{\rN}{\ensuremath{\underline{r}_\mathrm{N}}}
\newcommand{\RN}{\ensuremath{R_\mathrm{N}}}

\newcommand{\overlap}{\ensuremath{\varepsilon}}

\newcommand{\MSDmon}{\ensuremath{h}}
\newcommand{\MSDcmN}{\ensuremath{h_\mathrm{N}}}

\newcommand{\DN}{\ensuremath{D_\mathrm{N}}}

\newcommand{\TN}{\ensuremath{T_\mathrm{N}}}

\newcommand{\CN}{\ensuremath{C_\mathrm{N}}}

\newcommand{\tstar}{\ensuremath{t^*}}

\newcommand{\deltat}{\ensuremath{\delta t}}

\begin{document}

\title{Note: Scale-free center-of-mass displacement correlations in polymer\\
       films without topological constraints and momentum conservation}

\author{J.P.~Wittmer}
\email{joachim.wittmer@ics-cnrs.unistra.fr}
\affiliation{Institut Charles Sadron, Universit\'e de Strasbourg, CNRS, 23 rue du Loess, 67034 Strasbourg Cedex, France}
\author{N.~Schulmann}
\affiliation{Institut Charles Sadron, Universit\'e de Strasbourg, CNRS, 23 rue du Loess, 67034 Strasbourg Cedex, France}
\author{P.~Poli\'nska}
\affiliation{Institut Charles Sadron, Universit\'e de Strasbourg, CNRS, 23 rue du Loess, 67034 Strasbourg Cedex, France}
\author{J. Baschnagel}
\affiliation{Institut Charles Sadron, Universit\'e de Strasbourg, CNRS, 23 rue du Loess, 67034 Strasbourg Cedex, France}
\date{\today}
\maketitle

Polymer melts without topological constraints are believed to be described by the Rouse model 
\cite{DoiEdwardsBook}, i.e. the random forces from the bath acting on the center-of-mass (CM) $\rN(t)$ 
of a reference chain of length $N$ are supposed to be uncorrelated. For the corresponding 
CM mean-square displacement (MSD) $\MSDcmN(t)$ this implies 
\begin{equation}
\MSDcmN(t) \equiv \la (\rN(t) - \rN(0))^2 \ra =  2d\DN t
\label{eq_MSDcmNdef}
\end{equation}
for all times $t$ with $d$ being the spatial dimension, $\DN \approx b^2 W/N$ the self-diffusion coefficient and
$b$ the effective bond length \cite{DoiEdwardsBook}. The monomer mobility $W$ may be obtained by fitting the monomer MSD
\begin{equation}
\MSDmon(t) \equiv \la (\rvec(t) - \rvec(0))^2 \ra = b^2 (W t)^{2\alpha} 
\mbox{ with  } \alpha = 1/4
\label{eq_MSDmon}
\end{equation}
($\rvec(t)$ being the monomer position)
for times smaller than the relaxation time $\TN \approx N^2/W$ \cite{DoiEdwardsBook}.
In fact, correlated (colored) forces have been observed both experimentally and numerically 
as reviewed in Ref.~\cite{PaulGlenn}. These colored forces are best captured in a computer 
simulation using the ``velocity correlation function" (VCF) \cite{WPC11}
\begin{equation}
\CN(t) \equiv
\la \frac{\uvec(t)}{\deltat} \cdot \frac{\uvec(0)}{\deltat} \ra
\approx \frac{1}{2} \frac{\partial^2 \MSDcmN(t)}{\partial t^2} \sim \deltat^0 \mbox{ for } t \gg \deltat,
\label{eq_CNdef}
\end{equation}
which measures directly the curvature of $\MSDcmN(t)$ with respect to time with
$\uvec(t)=\rN(t+\deltat)-\rN(t)$ being the CM displacement for a time window $\deltat$.
Using simple scaling arguments it has recently been suggested \cite{WPC11} that an interplay of 
melt incompressibility and chain connectivity implies a negative algebraic decay of the VCF
\begin{equation}
\CN(t) = - c \frac{\DN W}{b^d \rho} (W t)^{-\omega} \mbox{ for } t \ll \TN \approx N^2/W
\label{eq_VCFkey}
\end{equation}
with $c$ being an empirical amplitude, $\rho$ the monomer density and
$\omega = (2+d) \alpha$ the exponent characterizing the colored forces.
This scaling has been demon\-strated numerically by means of Monte Carlo (MC) simulations
for three-dimensional (3D) melts where $\omega=5/4$ \cite{WPC11}.
The aim of this {\em Note} is to show that Eq.~(\ref{eq_VCFkey}) holds also in effectively 
two-dimensional (2D) polymer films.

As in Ref.~\cite{WPC11} we use a version of the bond-fluctuation model (BFM) with finite 
excluded volume penalty $\overlap$ and topology violating local moves to the 26 neighboring lattice sites. 
Length scales are given in units of the lattice constant, time in Monte Carlo Steps (MCS).
The chains are confined to a slit of width $H=4$ between two parallel repulsive walls following Ref.~\cite{CMWJB05}. 
All properties discussed refer to their 2D projection parallel to the walls. 
Working at the 3D volume fraction $\phi=0.5$ of occupied lattice sites \cite{CMWJB05}, 
the relevant 2D number density is thus $\rho = (\phi/8) H = 0.25$. 
The systems are effectively $d=2$ dimensional since the width of the slit is much smaller than the typical chain size.
For the overlap penalty $\overlap = 10$ (in units of $\kBT$) presented here the static properties are similar to
Ref.~\cite{CMWJB05} where $\overlap=\infty$.
Note that the monomers can still cross each other in the direction perpendicular to the walls. 
The chains are thus ``self-avoiding trails" 
\cite{ANS03}, i.e.  they adopt logarithmically swollen configurations \cite{ANS03,CMWJB05}. 
Since $\RN^2/N$ diverges for large $N$ (albeit very weakly),
the operational definition of the effective 2D bond length $b$ is a delicate issue
and the projected effective bond length of the bulk,
$b \equiv 3.24 \sqrt{2/3} \approx 2.65$, is used as reference.
Note that the scaling of the dynamical properties discussed below does not depend on this specific value.

\begin{figure}[t]
\centerline{\resizebox{0.95\columnwidth}{!}{\includegraphics*{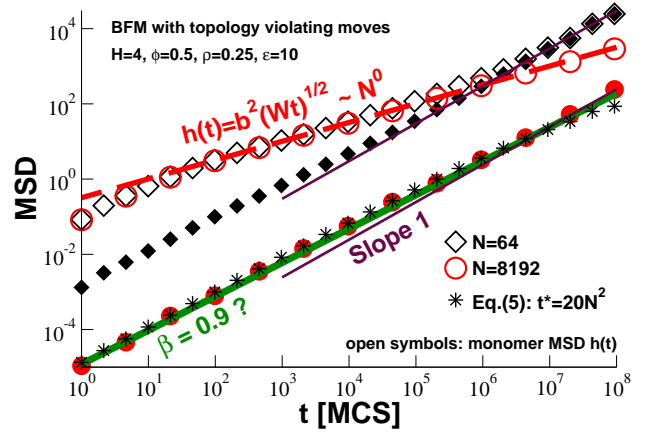}}}
\caption{Various MSDs for polymer melts confined to ultrathin films of width $H=4$
computed using the BFM algorithm with topology violating local MC moves \cite{WPC11}.
The open symbols represent the monomer MSD $\MSDmon(t)$ which compare well to the Rouse 
prediction (dashed line). Systematic deviations from the Rouse model are revealed
for the CM MSD $\MSDcmN(t)$ (filled symbols) as emphasized by the effective power-law 
exponent $\beta = 0.9$ (bold line). 
The stars correspond to Eq.~(\ref{eq_MSDcmNcorrect}) with $c=1$ and $\tstar =  20 N^2$ for $N=8192$.
\label{fig_MSD}
}
\end{figure}

It is important to stress that due to the use of finite overlap penalty and topology non-conserving MC moves, 
the dynamics remains {\em to leading order} of Rouse-type as can be seen from the MSDs presented in Fig.~\ref{fig_MSD}. 
The monomer MSD $\MSDmon(t)$ (open symbols) compares well over several orders of magnitude to Eq.~(\ref{eq_MSDmon}) 
indicated by the dashed line.
Although it is likely that this scaling does not hold in a strict sense due to the mentioned logarithmic 
swelling of the chains, deviations --- if they exist --- are apparently of higher order and irrelevant 
on the logarithmic scales we focus on.
As can be seen for $N=64$, the monomers diffuse again freely (thin solid line) for times $t \gg \TN$. 
Please note that it was not the aim of the present work to sample for our larger chains ($N > 2000$) over the
huge times needed to make this regime accessible. The short-time behavior,
Eq.~(\ref{eq_MSDmon}), can be used to determine $W \approx 0.002$ for the effective local
mobility and to predict a self-diffusion coefficient $\DN \approx 0.005/N$ using the 
Rouse model \cite{WPC11}. 
Systematic deviations from the Rouse scaling are revealed, however, for the CM MSD $\MSDcmN(t)$ (filled symbols)
as emphasized by the effective exponent $\beta = 0.9$ (bold line). 
At variance to the 3D bulk \cite{WPC11} deviations are visible up to a time $\tstar$ scaling 
as the chain relaxation time $\TN$.
(It is thus only possible to confirm the given value of $\DN$ from the long-time plateau of 
$\MSDcmN(t)/2dt$ for chains up to $N=1024$.)

\begin{figure}[t]
\centerline{\resizebox{0.95\columnwidth}{!}{\includegraphics*{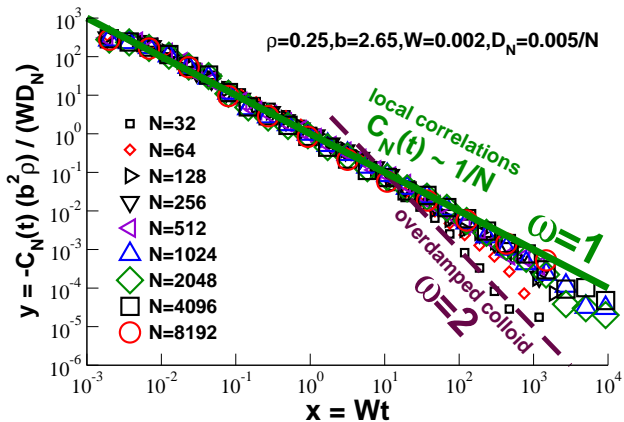}}}
\caption{Scaling plot of the directly computed displacement correlation function $\CN(t)$ for 
a large range of chain lengths $N$ (open symbols) using the parameters $\rho$, $b$, $W$ and $\DN \sim 1/N$ 
indicated in the plot. The data collapse for $t \ll \TN$ confirms that $\CN(t) \sim 1/N$.
The corresponding time exponent $\omega=1$ (bold line) is observed over up to five orders of magnitude in time. 
The exponent $\omega=2$ for overdamped colloids expected for $t \gg \TN$ is indicated by the dashed line.
\label{fig_VCF}
}
\end{figure}

A more precise characterization of the colored forces acting on the chains is
achieved by means of the VCF $\CN(t)$ computed directly 
as described in \cite{WPC11}. The rescaled data presented in Fig.~\ref{fig_VCF}
confirm the predicted algebraic decay, Eq.~(\ref{eq_VCFkey}), with an exponent $\omega=1$ and 
an empirical amplitude $c \approx 1$ (bold line). 
Note that the scaling implies $\CN(t) \sim 1/N$, i.e. the correlations are due to 
$\sim N$ local independent events as for 3D melts. 
The large time behavior for $t \gg \TN$ can be seen for the shorter chains which approach
the exponent $\omega = (2+d)/2 = 2$ (dashed line) expected for overdamped colloids
in incompressible solutions \cite{WPC11,DhontBook}.
Using Eq.~(\ref{eq_CNdef}) it follows from Eq.~(\ref{eq_VCFkey}) that
\begin{equation}
\MSDcmN(t) = 4 \DN t \left( 1 + \frac{c}{2 b^2 \rho} \ln(\tstar/t) \right)
\mbox{ for } t \ll \TN
\label{eq_MSDcmNcorrect}
\end{equation}
with $\tstar$ being an integration constant. 
Please note that the logarithmic contribution to Eq.~(\ref{eq_MSDcmNcorrect}) dominates for $t \ll \tstar$.
While in 3D we have $\tstar \sim N^0 \ll \TN$ \cite{WPC11}, it turns out that for thin films 
$\tstar \approx 20 N^2 \approx \TN/18$ yields a good fit for all chain lengths we have sampled
as can be seen for $N=8192$ in Fig.~\ref{fig_MSD} (stars). 

In summary, we have presented here computational work on the displacement
correlations in thin polymer films without topological constraints and
momentum conservation using a well-known lattice MC algorithm \cite{WCK09,WPC11}.
As anticipated by the scaling result suggested recently for 3D melts \cite{WPC11},
it is shown here for effectively 2D systems that the VCF $\CN(t)$ 
reveals a negative algebraic decay with a power-law exponent $\omega = (2+d)/4 = 1$.
This implies a logarithmic correction to the CM MSD.
Our MC approach corresponding to a strongly overdamped melt should be relevant to most
experimental setups due to the strong frictional forces from the walls.
Obviously, qualitatively different correlations are to be expected if hydrodynamic interactions matter 
--- 
as is the case for molecular dynamics simulations of a bead-spring model with a weak Langevin thermostat \cite{ANS11a}
---
or if the chains are confined to strictly 2D layers due to the ensuing compactness and surface fractality of the chains \cite{ANS03,WMJ10}. 


\end{document}